\documentclass{ws-ijmpd}

\begin{document}

\markboth{Lixin Xu, Hongya Liu} {Scaling Dark Energy in a
Five-Dimensional Bouncing Cosmological Model}
%
\catchline{}{}{}{}{}
%

\title{Scaling Dark Energy in a Five-Dimensional Bouncing
Cosmological Model}
\author{Lixin Xu\footnote{lxxu@student.dlut.edu.cn},
Hongya Liu\footnote{Corresponding author: hyliu@dlut.edu.cn}}
\address{Department of Physics, Dalian University of Technology,
Dalian, 116024, P. R. China}

\maketitle

\begin{history}
\received{Day Month Year} \revised{Day Month Year}
\end{history}

\begin{abstract}
We consider a 5-dimensional Ricci flat bouncing cosmological model
in which the 4-dimensional induced matter contains two components
at late times - the cold dark matter (CDM)+baryons and dark
energy. We find that the arbitrary function $f(z)$ contained in
the solution plays a similar role as the potential $V(\phi)$ in
quintessence and phantom dark energy models. To resolve the
coincidence problem, it is generally believed that there is a
scaling stage in the evolution of the universe. We analyze the
condition for this stage and show that a hyperbolic form of the
function $f(z)$ can work well in this property. We find that
during the scaling stage (before $z\approx 2$), the dark energy
behaves like (but not identical to) a cold dark matter with an
adiabatic sound speed $c_{s}^{2}\approx 0$ and $p_{x}\approx 0$.
After $z\approx 2$, the pressure of dark energy becomes negative.
The transition from deceleration to acceleration happens at
$z_{T}\approx 0.8$ which, as well as other predictions of the $5D$
model, agree with current observations.
\end{abstract}


\section{Introduction}

In recent decades, the observations of high redshift Type Ia
supernovae have revealed that our universe is undergoing an
accelerated expansion rather than decelerated expansion
\cite{RS,TKB,Riess}. Meanwhile, the discovery of Cosmic Microwave
Background (CMB) anisotropy on degree scales together with the
galaxy redshift surveys indicate $\Omega _{total}\simeq 1$
\cite{BHS} and $\Omega _{m}\simeq \left. 1\right/ 3$. All these
results strongly suggest that the universe is permeated smoothly
by 'dark energy' which has a negative pressure and violates the
strong energy condition. The dark energy and accelerating universe
have been discussed extensively from different points of view
\cite{Quintessence,Phantom,K-essence}.
In principle, a natural explanation to the cosmic acceleration is
the cosmological constant. However, there exist serious
theoretical problems such as the fine tuning problem and
coincidence problem. To overcome the fine tuning problem, a
self-interacting scaler field, dubbed quintessence, with an
equation of state (EOS) $w_{\phi}=p_{\phi}\left/
\rho_{\phi}\right.$ was introduced, where $w_{\phi}$ is time
varying and negative. Thus, by properly choosing the forms of the
potential $V\left(\phi\right)$, desired behaviors of the
quintessence can be obtained: (i) with a negative pressure which
drives the universe accelerated expansion; (ii) with an energy
density that was much smaller than that of the ordinary matter
(and radiation) at early times (due to the constrains from
primordial nucleosynthesis and structure formation) and is
comparable to the latter at recent times. To resolve the
coincidence problem, tracker solutions or scaling solutions were
designed in which there exists a stage where the ratio of
potential and kinetic energies of the scalar field remains
constant approximately. In this way, all the initial `information'
could be eliminated. So, in the past, if the ratio of the
densities of dark energy and matter is approximately constant, the
coincidence problem could be avoided.

It has been drawn great attention to the idea that our universe is
a 4-dimensional hypersurfce embedded in a higher dimensional world
as is in Kaluza-Klein theories and in brane world scenarios. Here
we consider the 5-dimensional Space-Time-Matter (STM) theory
\cite{Wesson}. This theory is distinguished from the classical
Kaluza-Klein theory by that it has an non-compact fifth dimension
and that it is empty viewed from $5D$ and sourceful viewed from
$4D$. That is, in STM theory, the $5D$ manifold is Ricci-flat
which implies that the five-dimensional space-time is empty, the
matter of our universe is induced from the fifth dimension, and
the $4D$ hypersurface is curved by this induced matter.
Mathematically, this approach is supported by Campbell's theorem
\cite{Compbell} which says that any analytical solution of
Einstein field equation of $N$ dimensions can be locally embedded
in a Ricci-flat manifold of $\left(N+1 \right)$ dimensions, though
there is an argument recently concerning Campbell's theorem
\cite{Anderson}.

In this paper we consider a class of five-dimensional cosmological
models \cite{LWX} of the STM theory. This class of exact solutions
satisfies the $5D$ Ricci-flat equations $R_{AB}=0$ and is
algebraically rich because it contains two arbitrary functions
$\mu \left( t\right) $ and $\nu \left( t\right)$. It was shown
\cite{LWX} that several properties characterize these $5D$ models:
(i) The $4D$ induced matter could be described by a perfect fluid
plus a variable cosmological 'constant'. (ii) By properly choosing
the two arbitrary functions, both the radiation-dominated and
matter-dominated standard FRW models could be recovered. (iii) The
big bang singularity of the $4D$ standard cosmology is replaced by
a big bounce at which the universe reaches to a finite and minimal
size. Before the bounce, the universe contracts; after the bounce,
it expands. Also, the evolution of the universe containing three
components was discussed in \cite{XL}. In this paper we assume
that the universe is permeated smoothly by CDM+baryons $\rho _{m}$
as well as dark energy $\rho _{x}$ with pressure
$p_{x}=w_{x}\rho_{x}$ at late time. We will show that a scaling
solution can be obtained which is different from the conventional
scaling solutions and late-time attractors: (i) There exists a
scaling stage before the transition from decelerated to
accelerated expansion, and so the coincidence problem could be
resolved; (ii) During this stage, the dark energy has the equation
of state $w_x\approx 0$ and behaves like a cold dark matter - this
would provide a unified model for dark matter and dark energy as
the Chaplygin gas model does \cite{unified}. Given an appropriate
form of $f(z)$, we will find that before $z\approx 2$, the dark
energy behaves like a `tracker' solution and mimics CDM+baryons.
The transition from deceleration to acceleration is at
$z_{T}\approx 0.8$, which agrees with recent cosmological
observations.

This paper is organized as follows. In section \ref{DE5}, we
assume that the induced matter of the universe contains two
components: CDM+baryons and dark energy. We write down Einstein
field equations and the equation of states (EOS) of dark energy.
In section \ref{P}, we rewrite these equations via redshift $z$,
then the arbitrary functions $\mu(t)$ contained in the $5D$
solution corresponds to a function $f\left(z\right)$, which is
found to play a similar role as the potential $V(\phi)$ in
quintessence and phantom models. With use of a kind of hyperbolic
function of $f\left(z\right)$, a scaling stage is obtained. Also,
we discuss the evolution of the dimensionless density parameters,
the EOS of dark energy, the deceleration parameter and the
clustering of dark energy. Section \ref{conclusion} is a
conclusion.

\section{Dark energy in the 5D model}\label{DE5}

Within the framework of STM theory, a class of exact $5D$
cosmological solution was presented by Liu and Mashhoon in 1995
\cite{Liu}. Then, in 2001, Liu and Wesson \cite{LWX} restudied the
solution and showed that it describes a cosmological model with a
big bounce as opposed to a big bang. The $5D$ metric of this
solution reads
\begin{equation}
dS^{2}=B^{2}dt^{2}-A^{2}\left( \frac{dr^{2}}{1-kr^{2}}+r^{2}d\Omega
^{2}\right) -dy^{2}  \label{5-metric}
\end{equation}
where $d\Omega ^{2}\equiv \left( d\theta ^{2}+\sin ^{2}\theta
d\phi ^{2}\right) $ and
\begin{eqnarray}
A^{2} &=&\left( \mu ^{2}+k\right) y^{2}+2\nu y+\frac{\nu ^{2}+K}{\mu ^{2}+k},
\nonumber \\
B &=&\frac{1}{\mu }\frac{\partial A}{\partial t}\equiv \frac{\dot{A}}{\mu }.
\label{A-B}
\end{eqnarray}
Here $\mu =\mu (t)$ and $\nu =\nu (t)$ are two arbitrary functions
of $t$, $k$ is the $3D$ curvature index $\left(k=\pm 1,0\right)$,
and $K$ is a constant. This solution satisfies the 5D vacuum
equation $R_{AB}=0$. So, the three invariants are
\begin{eqnarray}
I_{1} &\equiv &R=0, I_{2}\equiv R^{AB}R_{AB}=0,
\nonumber  \\
I_{3} &=&R_{ABCD}R^{ABCD}=\frac{72K^{2}}{A^{8}}. \label{3-invar}
\end{eqnarray}
The invariant $I_{3}$ in Eq. (\ref{3-invar}) shows that $K$
determines the curvature of the 5D manifold. Here we should
mention that because the $5D$ field equations of the STM theory
are always sourceless, i.e., $G_{AB}=0$ or $R_{AB}=0$, there is no
need to introduce a higher-dimensional Newtonian constant or a
higher-dimensional Planck mass in the theory. This is one of the
main differences between STM theory and other Kaluza-Klein
theories.

Using the $4D$ part of the $5D$ metric (\ref{5-metric}) to
calculate the $4D$ Einstein tensor, one obtains
\begin{eqnarray}
^{(4)}G_{0}^{0} &=&\frac{3\left( \mu ^{2}+k\right) }{A^{2}},
\nonumber \\
^{(4)}G_{1}^{1} &=&^{(4)}G_{2}^{2}=^{(4)}G_{3}^{3}=\frac{2\mu \dot{\mu}}{A%
\dot{A}}+\frac{\mu ^{2}+k}{A^{2}}.  \label{einstein}
\end{eqnarray}
In the previous work \cite{LWX}, the induced matter was set to be
a conventional matter plus a time variable cosmological
`constant'. In this paper, we assume that the induced matter
contains two parts: one is CDM+baryons $\rho _{m}$ with pressure
$p_{m}=0$, the other is the dark energy $\rho _{x} $ with pressure
$p_{x}$. So, we have
\begin{eqnarray}
\frac{3\left( \mu ^{2}+k\right) }{A^{2}} &=&\rho _{m}+\rho _{x},  \nonumber \\
\frac{2\mu \dot{\mu}}{A\dot{A}}+\frac{\mu ^{2}+k}{A^{2}}
&=&-p_{x}, \label{FRW-Eq}
\end{eqnarray}
where
\begin{equation}
p_{x}=w_{x}\rho _{x}.  \label{EOS-X}
\end{equation}
From Eqs.(\ref{FRW-Eq}) and (\ref{EOS-X}), one obtains the EOS of
dark energy
\begin{equation}
w_{x}=\frac{p_{x}}{\rho _{x}}=-\frac{2\left. \mu \dot{\mu}\right/
A \dot{A}+\left. \left( \mu ^{2}+k\right) \right/ A^{2}}{3\left.
\left( \mu ^{2}+k\right) \right/ A^{2}-\rho
_{m0}A^{-3}},\label{wx}
\end{equation}
and the dimensionless density parameters
\begin{eqnarray}
\Omega _{m} &=&\frac{\rho _{m}}{\rho _{m}+\rho _{x}}=\frac{\rho
_{m0}}{3\left( \mu ^{2}+k\right) A},  \label{omiga-M} \\
\Omega _{x} &=&1-\Omega _{m}.  \label{omiga-X}
\end{eqnarray}
where $\rho _{m0}=\bar{\rho}_{m0}A_{0}^{3}$. The proper
definitions of the Hubble and deceleration parameters should be
given as \cite{LWX},
\begin{eqnarray}
H&\equiv&\frac{\dot{A}}{A B}=\frac{\mu}{A} \\
q \left(t, y\right)&\equiv&\left.
-A\frac{d^{2}A}{d\tau^{2}}\right/\left(\frac{dA}{d\tau}\right)^{2}
=-\frac{A \dot{\mu}}{\mu \dot{A}}, \label{df}
\end{eqnarray}
from which we see that $\dot{\mu}\left/\mu\right.>0$ represents an
accelerating universe, $\dot{\mu}\left/\mu\right.<0$ represents a
decelerating universe. So the function $\mu(t)$ plays a crucial
role in determining the properties of the universe at late times.
It was pointed out that another arbitrary function
$\nu\left(t\right)$ may relate closely to the early epoch of the
universe \cite{XL}.

\section{Late time evolution of the cosmological parameters versus redshift}\label{P}

We consider the spatial flat case $k=0$. From the solution
(\ref{A-B}) we see that on a given $y=constant$ hypersurface we
have $A=A(t)$. So, without loss of generality, one can always
write $\mu=\mu(z)$ and $\nu=\nu(z)$ if $y$ is fixed, where $z$ is
the redshift, $A_{0}\left/A \right.=1+z$. Now we define
$\mu_{E}^{2}\left/ \mu_{z}^{2}\right.=f\left(z\right)$ and then we
find that Eqs. (\ref{wx})-(\ref{df}) reduce to
\begin{eqnarray}
w_{x} &=&-\frac{1+\left(1+z\right)d\ln
f\left(z\right)\left/dz\right.}{3-3\Omega_{m0}
\left(1+z\right)f\left(z\right)\left/f\left(0\right)\right.}, \label{wx-2} \\
\Omega _{m}
&=&\Omega_{mz_{E}}\frac{f\left(z\right)\left(1+z\right)}{1+z_{E}}=
\Omega_{m0}\frac{\left(1+z\right)f\left(z\right)}{f\left(0\right)},
\label{omigam-2} \\
\Omega _{x} &=&1-\Omega_{mz_{E}}\frac{f\left(z\right)\left(1+z\right)}{1+z_{E}},  \label{omigax-2} \\
q&=&\frac{1+3\Omega_{x}w_{x}}{2}=-\frac{\left(1+z\right)}{2}\frac{d\ln
f\left(z\right)}{dz}. \label{q}
\end{eqnarray}
Here the subscript $E$ denotes the point where
$\Omega_{mz_{E}}=\Omega_{xz_{E}}$ (which is $1/2$ when the
radiation is neglected). As is known from quintessence and phantom
dynamical dark energy models, the potential $V\left( \phi \right)$
is unfixed. By choose different forms of $V\left(\phi\right)$,
desired properties of dark energy could be obtained \cite{sahni}.
Now, there is an arbitrary function $f\left(z\right)$ in the
present $5D$ model. Different choice of $f\left(z\right)$
corresponds to different choice of the potential
$V\left(\phi\right)$. This enables us to look for desired
properties of the universe via Eqs. (\ref{wx-2})-(\ref{q}).

To obtain a scaling solution, there should be a stage where the
ratio of the energy densities of dark energy and dark matter is
approximately constant. So, the condition for exiting a scaling
stage at late time is $\frac{\Omega_{m}}{\Omega_{x}}\sim const$.
From Eqs. (\ref{omigam-2}) and (\ref{omigax-2}), this condition
leads to
\begin{equation}
\frac{d\left[f\left(z\right)\left(1+z\right)\right]}{dz} \approx
0, \label{condition}.
\end{equation}
Also, to obtain an accelerating universe at late times, some
mechanics is needed to break the scaling state steeply and make
the dark energy dominated. In \cite{AC}, a hyperbolic function
$k(\phi) = k_{min} + tanh(\phi-\phi_{1}) + 1$ as a leap kinetic
term to achieve the desired features of quintessence is discussed.
So, in our case, to have a scaling dark energy, we let $f(z)$
being of a similar form,
\begin{equation}
f\left(z\right)
=\frac{\tanh\left[a\left(z-b\right)\right]+c}{1+z}, \label{f}
\end{equation}
where, $a$, $b$ and $c$ are arbitrary constants and can be
estimated by observation data, i.e. by using
$\Omega_{mz_{E}}=0.5$, $\Omega_{m0}=0.273_{-0.041}^{+0.042}$,
$q_{0}= -0.67\pm0.25$. Then we obtain the values of $z_{E}$, $a$
and $b$ as
\begin{equation}
a=\frac{1}{2z_{E}}ln\frac{\left(1+\delta\right)\left(1+\Delta\right)}{\left(1-\delta\right)\left(1-\Delta\right)},\quad
b=\frac{z_{E}ln\frac{1+\Delta}{1-\Delta}}
{ln\frac{\left(1+\delta\right)\left(1+\Delta\right)}{\left(1-\delta\right)\left(1-\Delta\right)}},
\label{para}
\end{equation}
where, $\delta=1+z_E-c$ and
$\Delta=c-(1+z_E)\frac{\Omega_{m0}}{\Omega_{mz_{E}}}$. Using the
equation $q_{0}=-\frac{1}{2}\left.\frac{ d\ln
f\left(z\right)}{dz}\right|_{z=0}$, we can solve $z_{E}$
numerically by leaving $c$ as a free parameter which could be
determined by the constraints from structure formation (In this
paper we take $c=1.45$). So, the function $f\left(z\right)$
describes the evolution of CDM+baryons and dark energy completely
at late time.

\begin{figure}
\begin{center}
\epsfbox{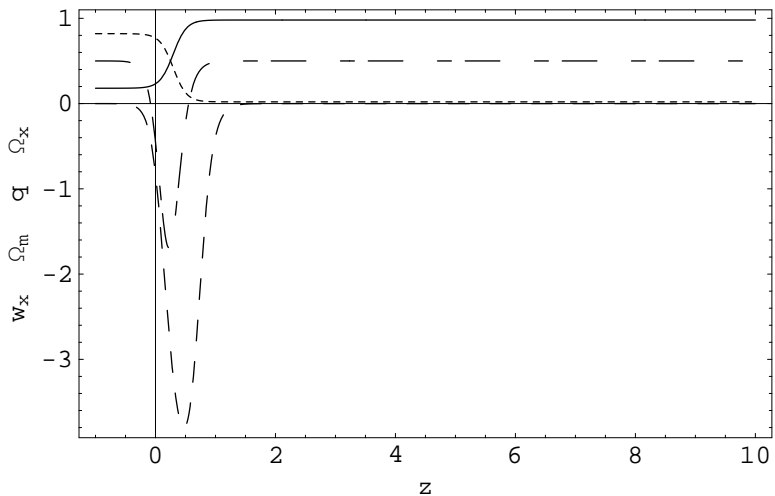}
\end{center}
\caption{The dimensionless density parameters  versus redshift
$z$: $\Omega_{x}$ of dark energy (dotted line, the third line from
the top), $\Omega_{m}$ of CDM+baryons (solid line, the first line
from the top), EOS of dark energy $w_{x}$ (dashed line, the forth
line from the top), and the deceleration parameter $q$
(dash-dotted line, the second line from the top). The figure is
plotted with $a=4.52$, $b=0.29$ and $c=1.45$. In these values, the
current values of the cosmological parameters are
$\Omega_{m0}=0.232$, $\Omega_{x0}=0.768$, $w_{x0}=-0.82$ and
$q_{0}=-0.45$, respectively. From $z=10$ to $z=400$, we find
$\Omega_x / \Omega_m\approx 0.02$ which is indeed a scaling.}
\label{all}
\end{figure}

\begin{figure}
\begin{center}
\epsfbox{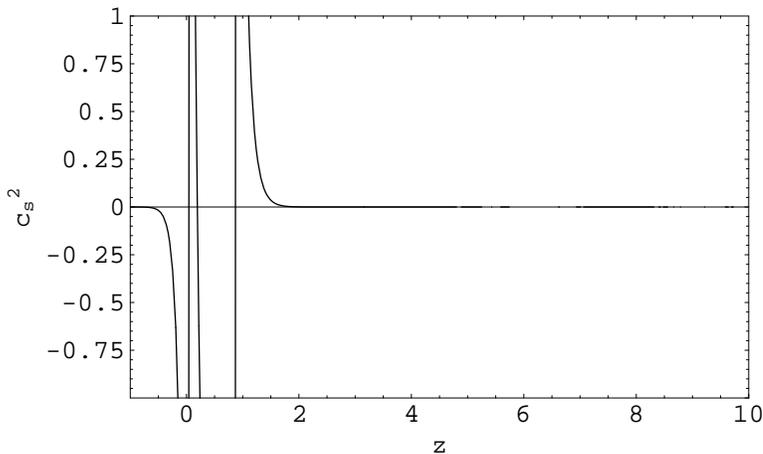}
\end{center}
\caption{The evolution of the adiabatic sound speed versus the
redshift $z$. The figure is plotted with $a=4.52$, $b=0.29$ and
$c=1.45$. In these values, the current values of the cosmological
parameters are $\Omega_{m0}=0.232$, $\Omega_{x0}=0.768$,
$w_{x0}=-0.82$ and $q_{0}=-0.45$, respectively.} \label{fs}
\end{figure}

\begin{figure}
\begin{center}
\epsfbox{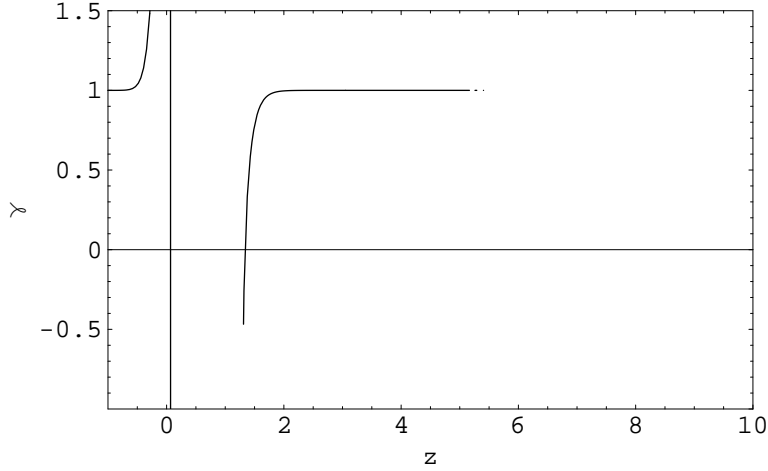}
\end{center}
\caption{The evolution of $\gamma$ versus redshift $z$. In the
scaling stage, $f \approx 1$ behaves like dark matter. In the
breaking stage, three singular points exist. The figure is plotted
with $a=4.52$, $b=0.29$ and $c=1.45$. In these values, the current
values of the cosmological parameters are $\Omega_{m0}=0.232$,
$\Omega_{x0}=0.768$, $w_{x0}=-0.82$ and $q_{0}=-0.45$,
respectively.}\label{gamma}
\end{figure}

Under the choice of the function $f\left(z\right)$ in the form of
(\ref{f}), the ratio of the dimensionless density parameters
$\Omega_m$ and $\Omega_x$ becomes constant when $z>z_{s}$, that is
to say, the dark energy is scaling with the matter.
And the EOS of dark energy is $w_{x}\approx 0$, though it is
negative in nature. Thus we conclude that just not a long time
ago, the EOS became negative and the dark energy began to drive
the universe to accelerate. For different values of $\Omega_{m0}$
and $q_{0}$, the parameters $z_{T}$, $z_{E}$ and $w_{x0}$ are
listed in Table \ref{T-1}.
\begin{table}[tbp]
\begin{center}
\begin{tabular}{cllllllll}
\hline Parameters & \multicolumn{8}{c}{For $\Omega_{m0}=.232$}
\\\hline
$q_{0}$       & -.25 & -.35 &  -.45 &  -.55 & -.65 & -.75 & -.85 & -.95 \\
$z_{E}$       & 1.03 &  .82 &  .71 &  .63 &  .57 &  .52 &  .48 &  .45    \\
$z_{T}$       &  .06 &  .00 &  .95 &  .94 &  .92 &  .89 &  .86 &  .83 \\
$w_{x0}$      & -.65 & -.74 & -.82 & -.91 & -1.00 & -1.09 & -1.17 & -1.26 \\
\hline  Parameters & \multicolumn{8}{c}{For $\Omega_{m0}=.273$}
\\\hline
$q_{0}$       & -.25 & -.35 &  -.45 &  -.55 & -.65  & -.75  & -.85  & -.95 \\
$z_{E}$       &  .71 &  .61 &  .54 &  .48 &  .44 &  .40 &  .37 &  .35 \\
$z_{T}$       &  .06 &  .88 &  .88 &  .87 &  .84 &  .82 &  .79 &  .76 \\
$w_{x0}$      & -.69 & -.78 & -.87 & -.96 & -1.05 & -1.15 & -1.24 & -1.33  \\
\hline  Parameters & \multicolumn{8}{c}{For $\Omega_{m0}=.314$}
\\\hline
$q_{0}$       & -.25 & -.35 &  -.45 &  -.55 & -.65 & -.75 & -.85 & -.95 \\
$z_{E}$       &  .52 &  .45 &  .40 &  .36 &  .33 &  .31 &  .28 &  .26  \\
$z_{T}$       &  .06 &  .83 &  .82 &  .81 &  .78 &  .76 &  .73 &  .71 \\
$w_{x0}$      & -.73 & -.83 & -.92 & -1.02 & -1.12 & -1.21 & -1.31 & -1.41 \\
\hline
\end{tabular}
\caption{Values of parameters $z_{T}$, $z_{E}$ and $w_{x0}$ for
different $\Omega_{m0}$ and $q_{0}$ and for $c=1.45$.}\label{T-1}
\end{center}
\end{table}
We can see that the transition point from deceleration to
acceleration is at $z_{T}\approx 0.8$, which is compatible with
current observations. Also, for a larger $\Omega_{m0}$, a larger
absolute value of $w_{x0}$ is needed to drive the universe
accelerating. For the case of $\Omega_{m0}=0.232$, $q_{0}=-0.45$
(which corresponds to $a=4.52$ and $b=0.29$), we plot the
evolution of dimensionless density parameters $\Omega_{m}$,
$\Omega_{x}$, EOS of dark energy $w_{x}$, and deceleration
parameter $q$ versus redshift $z$ in Fig. \ref{all}.
The adiabatic sound speed of the dark energy is
\begin{equation}
c_{s}^{2}=w_{x}^{2}\frac{\partial p_{x}}{\partial
z}\left/\left({w_{x}\frac{\partial p_{x}}{\partial
z}-p_{x}\frac{\partial w_{x}}{\partial z}}\right)\right..
\end{equation}
where, in terms of the redshift $z$, the pressure $p_x$ is
\begin{eqnarray}
p_x&=&-\frac{\mu^2_E(1+z)^2}{A^2_0 f(z)}\left(\frac{(1+z)d\ln
f(z)}{dz}+1\right) \nonumber \\
&=&H_0^2f(0)\frac{(1+z)^2}{f(z)}\left(\frac{(1+z)d\ln
f(z)}{dz}+1\right).
\end{eqnarray}
We plot the adiabatic sound speed $c_s^2$ in Fig. \ref{fs}, where
$a=4.52$, $b=0.29$ and $c=1.45$.
From Fig. \ref{fs} we find that the adiabatic sound speed of the
dark energy is $c_{s}^{2}\approx 0$ during the scaling stage. So,
it behaves like the conventional matter with an almost zero
$w_{x}$. However, for $z<2$, the sound speed $c_{s}^{2}$ becomes
wild and even negative - this is also encountered in other unified
dark matter and dark energy models.
Note that before $z=2$ the dark energy with $c^{2}_{s}\approx 0$
and $w_{x} \approx 0$ behaves like a dark matter (but not
identical to it because $w_{x}$ is negative in nature). However,
one should take into account the clustering of dark energy
seriously. As did in Ref. \cite{DCS}, we consider a universe
containing only the dark energy with $c^{2}_{s}\approx 0$ and the
perturbation $\delta=\delta \rho_{x}/\rho_{x}$, where $\rho_{x}$
is the dark energy density. Since $w_{x}\ne 0$, the Jean's
instability equation is modified,
\begin{equation}
a^{2}\frac{d^{2}\delta}{d a^{2}}+\frac{3}{2}aA[c^{2}_{s},
w_{x}]\frac{d\delta}{d
a}+\left(\frac{\kappa^{2}c^{2}_{s}}{\mathcal{H}^{2}}-\frac{3}{2}B[c^{2}_{s},w_{x}]\right)\delta=0,\label{perturb}
\end{equation}
where $a$ is the FRW scale factor,
$\mathcal{H}=\frac{a^{\prime}}{a}$ (the prime denotes the
derivative w.r.t. the conformal time),
$A[c^{2}_{s},w_{x}]=1-5w_{x}+2c^{2}_{s}$ and
$B[c^{2}_{s},w_{x}]=1-6c^{2}_{s}+8w_{x}-3w_{x}^{2}$. In general,
the ``growing'' solution is $\delta \propto a^{\gamma}$ where
$\gamma=\frac{1}{2}\left(1-\frac{3}{2}A+[(1-\frac{3}{2}A)^{2}+6B]^{1/2}\right)$.
For $c^{2}_{s}=w_{x}=0$, $\gamma=1$, the solution $\delta \propto
a$ is the conventional dust-dominated growing solution. For a
small but none-zero $c^{2}_{s}$ and $w_{x}$, the dark energy may
cluster during scaling. However, the process is slower than the
conventional dust-dominated case. And once $w_{x}< -0.12$ (for all
$c^{2}_{s} \ge 0$), $\gamma$ will less than zero and the dark
energy stop clustering. As \cite{DSW} does, the growth exponent
$f$ is defined as
\begin{equation}
f(a)\equiv \frac{d\ln \delta_k(a)}{d\ln a},
\end{equation}
which is roughly $k$-independent for a wide range of $k$. And, the
{\it growth factor} $g$ of density perturbations between arbitrary
$a_1 < a_2$ is defined as
\begin{equation}
g(a_1,a_2)\equiv
\frac{\delta_k(a_2)}{\delta_k(a_1)}=\left(\frac{a_2}{a_1}\right)^{\bar{f}}
\end{equation}
with a suitably defined average growth exponent $f$. For example,
the scaling stage begins at the dark matter dominated point
$a_{dec}$ (or $z_{dec}=1100$, decoupled epoch) and ends at
$a_{es}$ (or $z=2$), we have the growth factor
$g(a_{dec},a_{es})\approx 1100$. After the scaling stage and if
there is no clustering as expected, we have $g(a_{es},a_0)\approx
z_{es}^{\bar{f}}$ and
$\bar{f}=-\log_{z_{es}}(z_{dec}-z_{es})=-\log_{2}(1097)=-10$
roughly. In our case, the evolution of $\gamma$ versus redshift
$z$ is plotted in Fig. \ref{gamma} numerically.
In Fig. \ref{gamma}, we find that the singular points exist which
correspond to a negative infinity for $\gamma$. So, after this
stage, we cannot find any clustering dark energy in the universe
at present. However, we must point out that the wild properties of
the dark energy might be out of our knowledge and make the dark
energy more mysterious.

The influence of early dark energy on CMB anisotropy was discussed
in \cite{CDMSW}. It was pointed out that large $\Omega_x$ is ruled
out by cosmic observations. In our case, the dimensionless energy
density $\Omega_x$ in scaling stage depends on the parameter $c$
mainly, we can choose the parameters properly to fit observational
constraints. From Fig. \ref{all}, we find that $\Omega_{x}\approx
0.02$ at scaling stage which is in compatible with observations.
In \cite{DSW}, a general formula to calculate $\sigma_8$ was given
as
\begin{equation}
\frac{\sigma_8(A)}{\sigma_8(B)}\approx
(\frac{a_{tr}}{a_{eq}})^{3(\bar{\Omega_d^{sf}}(A)-\bar{\Omega_d^{sf}}(B))/5}
(1-\Omega_d^0)^{(\bar{\omega_B^{-1}}-\bar{\omega_A^{-1}})/5}\sqrt{\frac{\tau_0(A)}{\tau_0(B)}}.
\end{equation}
In our case, during the scaling stage, the dark energy behaves
like dark matter with a zero pressure. So, comparing the model
with the standard cold dark matter (SCDM) ($\sigma_8(SCDM)\approx
0.5\pm 0.1$), we have
\begin{equation}
\sigma_8(x)=1100^{-3(0.02-0)/5}\sigma_8(SCDM)=0.37\sigma_8(SCDM)\approx0.46.
\end{equation}
Moreover, observations such as cluster abundance constraints yield
\cite{WS}
\begin{equation}
\sigma_8=(0.5\pm0.1)\Omega_m^{-\gamma}
\end{equation}
where $\gamma$ is slightly model dependent and usually
$\gamma\approx 0.5$. So, in our case, we have
\begin{equation}
\sigma_8^{clus.}(x)=0.5\times0.98^{-0.5}\approx0.51.
\end{equation}
So, the ratio is
\begin{equation}
\frac{\sigma_8(x)}{\sigma_8^{clus.}(x)}\approx0.92,
\end{equation}
which is close to unity and compatible with observations.

\section{Conclusion}\label{conclusion}

A general class of $5D$ cosmological models is characterized by a
big bounce as opposed to the big bang in $4D$ standard
cosmological model. This exact solution contains two arbitrary
functions $\mu(t)$ and $\nu(t)$, which are in analogy with the
different forms of the potential $V\left(\phi\right)$ in
quintessence or phantom dark energy models. Also, once the forms
of the arbitrary functions are specified, the universe evolution
will be determined. In this bounce model, we assume the universe
to contain at late time two components: CDM+baryons and dark
energy. Instead of choosing the forms of the arbitrary functions
$\mu\left(t\right)$ and $\nu\left(t\right)$, we transform
$\mu\left(t\right)$ from coordinate time $t$ to redshift $z$. So,
the choice of $\mu\left(t\right)$ becomes the choice of
$f\left(z\right)$. Then, we let $f\left(z\right)$ to be of the
form of a hyperbolic function (\ref{f}) and study the evolution of
the universe. We find that there is a scaling stage before
$z\approx 2$, in which the dark energy has constant ratio with
CDM+baryons and behaves like a pressureless cold dark matter with
$w_{x}\approx 0$. After that, the pressure of the dark energy
deviates from $w_{x}\approx 0$ and gradually becomes much more
negative. The transition from decelerated expansion to accelerated
expansion is at $z_{T}\approx 0.8$ which, together with other
predictions of the present $5D$ model, is in agreement with
current observations. Thus we see that the arbitrary function
$f\left(z\right)$ of the $5D$ model plays a similar role as the
potential $V\left(\phi\right)$ in quintessence or phantom models
in eliminating the initial `information', then the coincidence
problem is resolved. Also, this kind of model would provide a
unified model for dark matter and dark energy.

However, we also see from Fig \ref{fs} and \ref{gamma} that
several singular points for $z <2$. Perhaps, these singularities
correspond to some unknown phase transition.

\section*{Acknowledgments}This work was supported by NSF (10273004), (10573003) and NBRP
(2003CB716300) of P. R. China.


\begin{thebibliography}{99}

\bibitem[1]{RS} A.G. Riess, et.al., {\it Observational evidence from supernovae for an
accelerating universe and a cosmological constant}, 1998 {\it
Astron. J.} {\bf 116} 1009, astro-ph/9805201; S. Perlmutter,
et.al., {\it Measurements of omega and lambda from 42
high-redshift supernovae}, 1999 {\it Astrophys. J.} {\bf 517} 565,
astro-ph/9812133.

\bibitem[2]{TKB} J.L. Tonry, et.al., {\it Cosmological Results from High-z Supernovae
}, 2003 {\it Astrophys. J.} {\bf 594} 1, astro-ph/0305008; R.A.
Knop, et.al., {\it New Constraints on $\Omega_M$,
$\Omega_\Lambda$, and w from an Independent Set of Eleven
High-Redshift Supernovae Observed with HST}, astro-ph/0309368;
B.J. Barris, et.al., {\it 23 High Redshift Supernovae from the IfA
Deep Survey: Doubling the SN Sample at z>0.7}, 2004 {\it
Astrophys.J.} {\bf 602} 571, astro-ph/0310843.

\bibitem[3]{Riess} A.G. Riess, et.al., {\it Type Ia Supernova Discoveries
at $z>1$ From the Hubble Space Telescope: Evidence for Past
Deceleration and Constraints on Dark Energy Evolution},
astro-ph/0402512.

\bibitem[4]{BHS} P. de Bernardis, et.al., {\it A Flat Universe from High-Resolution
Maps of the Cosmic Microwave Background Radiation}, 2000 {\it
Nature} {\bf 404} 955, astro-ph/0004404; S. Hanany, et.al., {\it
MAXIMA-1: A Measurement of the Cosmic Microwave Background
Anisotropy on angular scales of 10 arcminutes to 5 degrees},2000
{\it Astrophys. J.} {\bf 545} L5, astro-ph/0005123; D.N. Spergel
et.al., {\it First Year Wilkinson Microwave Anisotropy Probe
(WMAP) Observations: Determination of Cosmological
Parameters},2003 {\it Astrophys. J.} Supp. {\bf 148} 175,
astro-ph/0302209.

\bibitem[5]{Quintessence} I. Zlatev, L. Wang, and P.J. Steinhardt ,
\textit{Quintessence, Cosmic Coincidence, and the Cosmological
Constant}, 1999 {\it Phys. Rev. Lett.} {\bf 82} 896,
astro-ph/9807002; P.J. Steinhardt, L. Wang , I. Zlatev, {\it
Cosmological Tracking Solutions}, 1999 {\it Phys. Rev.} D {\bf 59}
123504, astro-ph/9812313; M.S. Turner , {\it Making Sense Of The
New Cosmology}, 2002 {\it Int. J. Mod. Phys.} A {\bf 17S1} 180,
astro-ph/0202008; V. Sahni , {\it The Cosmological Constant
Problem and Quintessence}, 2002, {\it Class.Quant.Grav.} {\bf 19}
3435, astro-ph/0202076.

\bibitem[6]{Phantom} R.R. Caldwell, M. Kamionkowski,
N.N. Weinberg, {\it Phantom Energy: Dark Energy with w $<-1$
Causes a Cosmic Doomsday}, 2003 {\it Phys. Rev. Lett.} {\bf 91}
071301, astro-ph/0302506; R.R. Caldwell , {\it A Phantom Menace?
Cosmological consequences of a dark energy component with
super-negative equation of state}, 2002 {\it Phys. Lett.} B {\bf
545} 23, astro-ph/9908168; P. Singh, M. Sami, N. Dadhich, {\it
Cosmological dynamics of a phantom field}, 2003 {\it Phys. Rev.} D
{\bf 68} 023522, hep-th/0305110; J.G. Hao, X.Z. Li , {\it
Attractor Solution of Phantom Field}, 2003 {\it Phys.Rev.} D {\bf
67} 107303, gr-qc/0302100.

\bibitem[7]{K-essence} Armend\'{a}riz-Pic\'{o}n, T. Damour, V. Mukhanov,
{\it k-Inflation}, 1999 {\it Physics Letters} B {\bf 458} 209; M.
Malquarti, E.J. Copeland , A.R. Liddle, M. Trodden, {\it A new
view of k-essence}, 2003 {\it Phys. Rev.} D {\bf 67} 123503; T.
Chiba , {\it Tracking k-essence}, 2002 {\it Phys. Rev.} D {\bf 66}
063514, astro-ph/0206298.

\bibitem[9]{LWX} H.Y. Liu and P.S. Wesson, {\it Universe models with a variable
cosmological ``constant'' and a ``big bounce''}, 2001 {\it
Astrophys. J.} {\bf 562} 1, gr-qc/0107093; T. Liko, P.S. Wesson,
gr-qc/0310067; S.S. Seahra, P.S. Wesson, {\it Universes encircling
five-dimensional black holes}, 2003 {\it J. Math. Phys.}, {\bf 44}
5664; Ponce de Leon J, 1988 {\it Gen. Relativ. Gravit.} {\bf 20}
539; L.X. Xu , H.Y. Liu, B.L. Wang, {\it Big Bounce singularity of
a simple five-dimensional cosmological model}, 2003 {\it Chin.
Phys. Lett.} {\bf 20} 995, gr-qc/0304049; H.Y. Liu, {\it Exact
global solutions of brane universe and big bounce}, 2003 {\it
Phys. Lett.} B {\bf 560} 149, hep-th/0206198; B.L. Wang, H.Y. Liu
, L.X. Xu , {\it Accelerating Universe in a Big Bounce Model},
2004 {\it Mod. Phys. Lett.} A {\bf 19} 449(2004), gr-qc/0304093.

\bibitem[10]{XL} L.X. Xu  and H.Y. Liu, {\it Three Components Evolution in
a Simple Big Bounce Cosmological Model}, astro-ph/0412241, to be
published in IJMPD.

\bibitem[11]{Wesson} P.S. Wesson {\it Space-Time-Matter} (Singapore: World
Scientific) 1999; J.M. Overduin and P.S. Wesson, \textit{Phys.
Rept.} \textbf{283}, 303 (1997), gr-qc/9805018.

\bibitem[12]{Compbell} J.E. Campbell, {\it A Course of Differential Geometry},
(Clarendon Oxford, 1926); S. Rippl, R. Romero, R. Tavakol, 1995
{\it Gen. Quantum Grav.} {\bf 12} 2411; C. Romero, R. Tavako and
R. Zalaletdinov, 1996 {\it Ge. Relativ. Gravit.} {\bf 28} 365; J.
E.Lidsey, C. Romero, R. Tavakol and S. Rippl, 1997 {\it Class.
Quantum Grav.} {\bf 14} 865; S.S. Seahra and P.S. Wesson, {\it
Application of the Campbell-Magaard theorem to higher-dimensional
physics}, 2003 {\it Class. Quant. Grav.} {\bf 20} 1321,
gr-qc/0302015.

\bibitem[13]{Anderson} E. Anderson, {\it The Campbell--Magaard Theorem is inadequate
and inappropriate as a protective theorem for relativistic field
equations}, gr-qc/0409122; F. Dahia and C. Romero, {\it
Dynamically generated embeddings of spacetime}, gr-qc0503103.

\bibitem[14]{Liu} H.Y. Liu and B. Mashhoon, {\it A machian interpretation of the
cosmological constant}, 1995 {\it Ann. Phys}. {\bf 4} 565.

\bibitem[15]{unified} F. Pedro Gonzalez-Diaz, {\it Unified Model for Dark
Energy}, astro-ph/0212414; V.F. Cardone, A. Troisi, S.
Capozziello, {\it Unified dark energy models : a phenomenological
approach}, Phys.Rev. D {\bf 69} (2004) 083517, astro-ph/0402228;
M.C. Bento, O. Bertolami, A.A. Sen, {\it The Revival of the
Unified Dark Energy-Dark Matter Model ?}, Phys.Rev. D {\bf 70}
(2004) 083519, astro-ph/0407239; Hongsu Kim, {\it Brans-Dicke
Theory as an Unified Model for Dark Matter - Dark Energy},
astro-ph/0408577; Zong-Hong Zhu, {\it Generalized Chaplygin gas as
a unified scenario of dark matter/energy: observational
constraints}, Astron.Astrophys. {\bf 423} (2004) 421,
astro-ph/0411039.

\bibitem[16]{Liu-b} H.Y. Liu, {\it Exact global solutions of brane universe and big bounce},
2003 {\it Phys. Lett.} B {\bf 560} 149, hep-th/0206198.

\bibitem[17]{sahni} V. Sahni, {\it Theoretical models of dark
energy}, 2003 {\it Chaos. Soli. Frac.} {\bf 16} 527.

\bibitem[18]{AC} A. Hebecker and C. Wetterich, {\it Natural
Quintessence ?}, 2001 {\it Phys. Lett.} B {\bf 497} 281.

\bibitem[19]{DSW} M. Doran, J.M. Schwindt, and C. Wetterich ,
{\it Structure formation and the time dependent of the
quintessence}, 2001 {\it Phys. Rev.} D {\bf 64} 123520.

\bibitem[20]{DCS} S. DeDeo, R.R. Caldwell, P.J. Steinhardt,
{\it Effects of the Sound Speed of the Quintessence on the
Microwave Background and Large Scale Structure}, {\it Phys. Rev.}
D {\bf 67} (2003) 103509, astro-ph/0301284; T. Padmanabhan, {\it
Structure Formation in the Universe}, Cambridge University Press
(1995).

\bibitem[21]{CDMSW} R.R. Caldwell, M. Doran, C.M. Mueller, G. Schaefer,
C. Wetterich, {\it Early Quintessence in Light of WMAP}, 2003 {\it
Astrophys.J.} {\bf 591} L75.

\bibitem[22]{WS} L. Wang and P.J. Steinhardt, {\it Astrophys. J.} {\bf
508}, 483 (1998).

\end{thebibliography}
\end{document}